\documentclass[superscriptaddress,aps,journal=prx,preprint]{revtex4-2}
\usepackage{graphicx}
\usepackage{epstopdf}
\usepackage{bm,amsmath,amssymb} 
\usepackage{wasysym}
\usepackage{color, soul} 
\usepackage{dcolumn}   
\usepackage{multirow}
\usepackage[colorlinks=true, linkcolor=black, citecolor=black, urlcolor=blue]{hyperref}
\newcommand{\etal}{{\em et al}.\ }

\newcommand{\eg}{{\em e.\,g.}}

\usepackage{lineno}
\begin{document}
\title{Bonding relay for room-temperature oxide plasticity like metals}
\author{Xiangkai Chen}
\thanks{These two authors contributed equally}
\affiliation{Department of Physics, School of Science, Westlake University, Hangzhou 310030, China}
\affiliation{Institute of Natural Sciences, Westlake Institute for Advanced Study, Hangzhou  310024, China}
\author{Yuhong Li}
\thanks{These two authors contributed equally}
\affiliation{Shenyang National Laboratory for Materials Science, Institute of Metal Research, Chinese Academy of Sciences, Shenyang 110016, China}
\affiliation{School of Materials Science and Engineering, University of Science and Technology of China, Shenyang 110016, China}
\author{Xiaofei Zhu}
\affiliation{Shenyang National Laboratory for Materials Science, Institute of Metal Research, Chinese Academy of Sciences, Shenyang 110016, China}
\author{Yun-Long Tang}
\email{yltang@imr.ac.cn}
\affiliation{Shenyang National Laboratory for Materials Science, Institute of Metal Research, Chinese Academy of Sciences, Shenyang 110016, China}
\affiliation{School of Materials Science and Engineering, University of Science and Technology of China, Shenyang 110016, China}
\author{Shi Liu}
\email{liushi@westlake.edu.cn}
\affiliation{Department of Physics, School of Science, Westlake University, Hangzhou 310030, China}
\affiliation{Institute of Natural Sciences, Westlake Institute for Advanced Study, Hangzhou 310024, China}

\begin{abstract}
Oxides have long been regarded as intrinsically brittle due to their strong, directional ionic or covalent bonds, in stark contrast to the ductile behavior of metals, where delocalized electron sharing enables plasticity through facile dislocation glide. 
Here, we challenge this paradigm by demonstrating that typical oxides, such as SrTiO$_3$ and MgO, can exhibit room-temperature plasticity with pronounced crystallographic anisotropy. 
Through an integrated approach combining \textit{ab initio} calculations, large-scale molecular dynamics simulations, and experimental nanoindentation, we identify a universal structural criterion enabling room-temperature oxide plasticity: the presence of alternating positively and negatively charged atomic layers along specific slip directions, specifically the $(1\bar{1}0)[110]$ orientation in perovskite and rocksalt oxides. 
This charge-alternating configuration enables a ``bonding relay'' mechanism, in which sequential bond breaking and reformation across the slip plane accompanied by interlayer persistent bonds mimics multi-centered interactions in metals, thereby facilitating dislocation motion without catastrophic failure.
Our findings reveal a previously unrecognized pathway to achieving metal-like plasticity in oxides and establish a structural design principle for engineering flexible and mechanically resilient oxide materials.

\end{abstract}

\maketitle
\newpage
\date{\today}

Metals exhibit good plasticity due to the multi-centered interactions between metal cations and delocalized electrons~\cite{C.S.Barrett92}. The balance between repulsive forces among positively charged cations and the electrostatic attraction with the surrounding electron cloud results in moderately strong yet flexible bonds. During deformation, these multi-centered interactions enable rapid reorganization of metallic bonds, preserving bonding integrity as atoms glide past one another~\cite{Smithells92}. Consequently, metal cations can slide without disrupting the overall structure, as illustrated in Fig.~1a, which facilitates efficient atomic slippage and underpins the ductile nature of metals. In contrast, oxides are typically brittle due to their strong directional ionic or covalent bonds~\cite{Green98}. These bonds resist atomic glide across crystal planes, making dislocation motion more difficult and requiring significantly higher Peierls stresses than in metals~\cite{Li22p6925, Porz21p1528, Li19p5519}. Even when atomic planes attempt to glide, the limited adaptability of the bonding network hinders rapid reconfiguration, often leading to lattice mismatch and structural failure. Consequently, oxides are prone to crack formation under mechanical stress. Given the wide range of functional properties exhibited by oxide materials in devices such as sensors~\cite{Trung16p4338}, memory devices~\cite{Schroeder22p653}, and electronics~\cite{Coll19p1}, developing ductile oxides with enhanced mechanical resilience could not only extend device lifespans but also unlock transformative applications in fields like flexible electronics and wearable sensors.

In recent years, single-crystal inorganic semiconductors such as Ag$_2$S, ZnS and Mg$_3$Bi$_2$ have been reported to exhibit metal-like plasticity, attributed to the presence of atomically easy glide planes~\cite{Shi18p421, Oshima18p772, Zhao24p777}. Intriguingly, single-crystal SrTiO$_3$, a prototypical perovskite oxide, has also displayed experimental signatures suggestive of plastic-like deformation~\cite{Brunner01p1161, Gumbsch01p085505, Yang09p2345, Fang24p81, Wang24p7442}. However, the underlying microscopic mechanisms remain elusive, and definitive evidence confirming intrinsic plasticity in single-crystal oxides is still lacking. Whether a universal design principle exists that enables traditionally brittle oxides to accommodate strain via dislocation activity and dislocation motion remains an open and compelling question.

In this work, we demonstrate through a combination of experimental and theoretical methods that oxides can exhibit plasticity along specific crystallographic orientations that enable a ``bonding relay” mechanism. As illustrated in Fig.~1b, this mechanism relies on the presence of alternating positively and negatively charged atomic layers, which facilitate sequential bond breaking and reformation during shear deformation. During lattice sliding, cations dynamically adjust their coordination with neighboring anions, allowing bonds to break and reform in a relay-like fashion. Throughout this process, each cation remains bonded to surrounding anions via at least one persistent bond, while the breaking of existing bonds is compensated by the simultaneous formation of new ones, referred to as relay bonds.  
This multi-centered interaction, reminiscent of metallic bonding, ensures continuous charge compensation across the sliding interface, effectively redistributing stress and maintaining structural integrity even under large shear strains. For representative oxides like MgO and SrTiO$_3$, we identify the (1$\Bar{1}$0)[110] crystallographic orientation as exhibiting the necessary structural prerequisites for bonding relay. Our large-scale molecular dynamics (MD) simulations, corroborated by experimental observations, confirm the presence of orientation-dependent (anisotropic) plasticity in these materials at room temperatures. 
Guided by this design principle, we predict a broad class of oxides, including CaO, SrO, and even ferroelectric PbTiO$_3$, that exhibit similar plastic behavior along lattice directions supporting bonding relay. These findings establish a framework for understanding and predicting plasticity in oxides and offer valuable design principles for the development of flexible oxide materials.

The atomistic structures of the (1$\bar{1}$0)[110] crystallographic orientation in MgO and SrTiO$_3$ are illustrated in Fig.~1e-f. Both materials with the (1$\bar{1}$0)[110] orientation exhibit alternating layers of positively charged cations (Mg$^{2+}$ and Sr$^{2+}$/Ti$^{4+}$) and negatively charged anions (O$^{2-}$), forming a charge-alternating configuration that is a key prerequisite for lattice sliding via the bonding relay mechanism. To evaluate the mechanical response of these oxides under shear, we perform density functional theory (DFT) calculations to construct shear stress--strain curves. As shown in Fig.~1c, both materials display an initial linear elastic regime, followed by a transition to plastic deformation at maximum shear stresses ($\sigma^*$) of 13.0~GPa for MgO and 14.3~GPa for SrTiO$_3$. These values exceed those of typical face-centered cubic (fcc) metals along the (111)[1$\bar{1}$0] crystallographic orientation such as Cu ($\sigma^*$~$=$~3.13~GPa), reflecting the higher stress required to nucleate dislocations in oxides.

To investigate the energetics of dislocation motion, we compute slip energy barriers for slip along the (1$\bar{1}$0)[110] crystallographic orientation, as shown in Fig.~1d. Notably, maximum slip energy barrier ($\gamma^*$), a key metric for slip resistance, is relatively low for both oxides (1.16~J/m$^2$ for MgO and 1.50~J/m$^2$ for SrTiO$_3$), comparable to values reported for fcc metals along the (111)[1$\bar{1}$0] crystallographic orientation ($\gamma$~$=$~0.5--1.0~J/m$^2$). Figure.~1g summarizes the values of $\sigma^*$ and $\gamma^*$ across a range of oxides and metals. While the $\sigma^*$ values for oxides (8--15~GPa) are higher than those of metals ($\leq$~4~GPa), their slip energy barriers are comparable.
This apparent decoupling between dislocation nucleation and glide suggests that, although oxides require significantly higher stress to initiate plasticity, once dislocations are nucleated, their subsequent motion along the [110] direction on (1$\bar{1}$0) planes proceeds with comparable energy efficiency to that in metals. These results challenge the prevailing perception of oxides as  intrinsically brittle and instead highlight their potential for metal-like plasticity under appropriate crystallographic orientations.

The role of charge-alternating layers in activating the bonding relay mechanism for plasticity is elucidated using SrTiO$_3$ as a representative system.
We note that Ti--O interactions are substantially stronger than Sr--O interactions (ICOHP values for Ti-O and Sr-O bonds are about 0.67 and 0.32, respectively), indicating that the shear behavior of SrTiO$_3$ along the (1$\bar{1}$0)[110] crystallographic orientation is predominantly governed by the Ti--O sublattice.
Figure~2a-c compares the atomic geometry of the Ti--O sublattice across three crystallographic orientations: ($1\bar{1}0$)[110], ($1\bar{1}0$)[001], and (001)[100]. Notably, the ($1\bar{1}0$)[110] orientation features well-defined alternating layers of Ti-only and O-only ions, an arrangement that allows for sequential rupture and reformation of bonds between anions and cations. In contrast, both the ($1\bar{1}0$)[001] and (001)[100] orientations exhibit mixed atomic layers containing both Ti and O ions.
These structural differences are reflected in the shear stress–strain behavior shown in Fig.~2d. The ($1\bar{1}0$)[110] orientation demonstrates a relatively low $\sigma^*$. In comparison, the (001)[100] orientation exhibits a much higher $\sigma^*$ of 40~GPa, indicating strong resistance to shear deformation. Notably, the ($1\bar{1}0$)[001] orientation shows no sign of plastic transformation.
Consistent with these trends, the slip energy barriers in Fig.~2e reveal that the ($1\bar{1}0$)[110] orientation has a much lower value of $\gamma^*$ compared to the other two orientations. Furthermore, Fig.~S3 shows that the cleavage energy ($\Delta E_c$) perpendicular to the ($1\bar{1}0$) plane is significantly higher than the slip energy barrier along the ($1\bar{1}0$) plane, suggesting that SrTiO$_3$ does not undergo cleavage during slip along the ($1\bar{1}0$) plane.
Together, these results highlight a direct link between atomic-scale layering and the potential for plastic deformation in oxides.

To quantify the bonding evolution during sliding (Figs.~2f–h), we analyze the evolution of Ti--O interactions in ($1\bar{1}0$)[110] SrTiO$_3$ by tracking the integrated crystal orbital Hamilton population (ICOHP) values for three categories of Ti--O bonds: persistent, broken, and relay bonds.
The magnitude of --ICOHP serves as a quantitative measure of bond strength.
Persistent bonds are Ti--O pairs that remain interacting throughout the glide process. Broken bonds are those that progressively weaken and ultimately rupture with increasing displacement. In contrast, relay bonds are newly formed interactions that emerge to restore local coordination, effectively maintaining the integrity of the bonding network.
As shown in Fig.~2i, the --ICOHP value of the presistant bond remains nearly constant. Importantly, the weakening of the broken bond is largely compensated by the concurrent strengthening of the relay bond. 
This dynamic compensation, where bond rupture is energetically offset by new bond formation, supports stable, continuous plastic deformation while preventing abrupt structural failure. The near-invariant total --ICOHP throughout the glide process (Fig.~2g) highlights the effectiveness of this bonding relay mechanism, which resembles the delocalized, multi-centered bonding interactions in ductile metals. For ($1\bar{1}0$)[001] direction, the absence of persistent bonds during the bonding relay leads to brittle fracture behavior (see Fig.~S2).
We perform a similar analysis for MgO, comparing the ($1\bar{1}0$)[110], ($1\bar{1}0$)[001], and (001)[100] slip directions (see Fig.~S3). The ($1\bar{1}0$)[110] orientation in MgO also displays clear signatures of plastic deformation, enabled by the presence of well-defined alternating cation and anion layers that support the bonding relay mechanism.

Taking SrTiO$_3$ and MgO as representative cases, the anisotropic plasticity of oxides is further confirmed through large-scale MD simulations, utilizing machine-learning-based interatomic force fields trained on a DFT database (see Supplementary Material for details). These simulations uncover the dynamic interplay between shear-induced dislocation activity and strain accommodation. Figure~3a–d  presents the microstructural evolution in bulk SrTiO\(_3\) under [110]-oriented shear deformation. At a shear strain of 19.2\%, we observe the nucealtion of paired partial dislocations (\eg, B$_1$ for $\vec{b} = -\frac{a}{2}[110]$ and B$_2$ for $\vec{b} = \frac{a}{2}[110]$ in Fig.~3a). As the applied shear strain increases to 20.0\%, these dislocations glide along the slip plane, transitioning to new positions (B$_1'$/B$_2'$ in Fig.~3b), while additional dislocation pairs (C$_1$/C$_2$ and E$_1$/E$_2$) nucleate to accommodate the increasing strain. 
Notably, dislocations such as D$_1$ and D$_2$ are also observed in the ($1\bar{1}0$) plane perpendicular to the primary shear direction, indicating the onset of three-dimensional dislocation network formation. These dislocation networks correspond to the slip bands produced in our nanoindentation experiments (see Fig.~4).
At 20.5\% shear strain (Fig.~3c), further loading drives interactions between mobile dislocations and other structural defects. For example, dislocation A$_2'$ merges with D$_1$, exemplifying a strain-hardening mechanism mediated by dislocation pinning. By the final strain state (Fig.~3d), an interconnected dislocation network forms through repeated cycles of dislocation nucleation, glide, and entanglement. This self-organized network redistributes localized stresses via dislocation-mediated slip and defect interactions, thereby enhancing ductility and mitigating catastrophic failure.
The sustained generation and mobility of dislocations under increasing shear strain underscore the critical role of the $(1\bar{1}0)[110]$ slip direction in enabling plastic deformation in SrTiO\(_3\). In contrast, MD simulations reveal that bulk SrTiO\(_3\) rapidly develops cracks under shear strain applied along the ($1\bar{1}0$)[001] and (001)[100] directions (see Fig.~S10 and S11).

To probe the dislocation glide mechanisms in SrTiO\(_3\), we simulate the dynamics of a pre-existing dislocation under [110]-oriented shear. As shown in Fig.~3e, equilibrium relaxation induces the dissociation of the initial dislocation into two partial dislocations ($\vec{b} = \frac{a}{2}[110]$), separated by an anti-phase domain boundary (APB), in agreement with the results shown in Fig.~3a and prior experimental observations of dislocation splitting in perovskites\cite{Castillo-Rodríguez10p270}. This dissociation is thermodynamically driven by the slip energy landscape (Fig.~1d): the characteristic double-hump profile along the \((1\bar{1}0)[110]\) slip pathway stabilizes a metastable APB between the partials, thereby favoring decomposition of the full dislocation ($\vec{b} = a[110]$) into two partials ($\vec{b} = \frac{a}{2}[110]$). Under applied shear strain (Fig.~3f–h), both partials glide cooperatively along the [110] direction while maintaining a constant separation distance, as dictated by the APB energy minimum. Notably, the invariant spacing between partials during slip reflects a dynamic equilibrium between shear-driven motion and APB-mediated elastic repulsion.
Our MD simulations are consistent with the experimental observations of Fang~\etal, who introduced dislocations into SrTiO\(_3\) through mechanical deformation. Their findings indicate that when grown-in dislocations are aligned with the glide plane, dislocation motion and multiplication can occur in SrTiO\(_3\) even at room temperature~\cite{Fang24p81}.

The microstructural evolution of bulk MgO under [110]-oriented shear deformation also display sustained dislocation activity, characterized by the continuous nucleation and glide of full dislocations ($\vec{b} = \frac{a}{2}[110]$) within ($1\bar{1}0$) planes (see Fig.~S12). Our MD simulations confirm that, unlike in SrTiO\(_3\), MgO retains unassociated full dislocations ($\vec{b} = \frac{a}{2}[110]$) (Fig.~3i). This stability is consistent with the slip energy barriers of MgO (Fig.~1d), which exhibits a single energy barrier and lacks the double-hump signature observed in SrTiO\(_3\). The absence of metastable minima eliminates the thermodynamic driving force for dislocation splitting. As a result, plasticity in MgO proceeds through the collective motion of full dislocations. For ($1\bar{1}0$)[001], despite [001]-aligned strain application, the dislocation glide occurs perpendicular to the ($1\bar{1}0$) plane (along [110] direction), yielding results similar to the ($1\bar{1}0$)[110] case (see Fig.~S13). The (001)[100] direction forms cracks rather than dislocations, with pre-existing dislocations being destabilized (Fig.~S14).
 
Finally, we validate the above simulation results through experimental nanoindentation on [100]-, [110]-, and [111]-oriented SrTiO\(_3\) single crystals (Fig.~4). The first pop-in event and subsequent multiple pop-in events provide clear evidence of plastic deformation in all these SrTiO\(_3\) samples (Figs.~4a, d, g). In particular, atomic force microscopy (AFM) surface topography and piezoresponse force microscopy (PFM) images reveal distinct slip features in the [100]-, [110]-, and [111]-oriented SrTiO\(_3\) single crystals.
For the (100), (110), and (111) surfaces, dislocation slip is observed along the [001] and [010] directions in [100]-oriented SrTiO\(_3\), along the \(\langle111\rangle\) and [001] directions in [110]-oriented SrTiO\(_3\), and along the \(\langle112\rangle\) directions in [111]-oriented SrTiO\(_3\). These slip characteristics correspond to dislocation motion along the \((1\bar{1}0)[110]\) slip direction via \(\langle110\rangle\)-type dislocations, as detailed in Fig.~S15.
Furthermore, nanoindentation experiments on [100]-, [110]-, and [111]-oriented MgO single crystals also exhibit clear evidence of plastic deformation and dislocation slip along similar crystallographic directions (Fig.~S16-17). These slip behaviors similarly correspond to \((1\bar{1}0)[110]\) dislocation motion via \( \frac{1}{2}a\langle110\rangle \)-type dislocations in the MgO crystal. 

In summary, we establish that a wide range of oxides that are traditionally considered brittle can exhibit intrinsic plasticity similar to metals when sheared along specific crystallographic orientations. Through a combined theoretical and experimental investigation of SrTiO$_3$ and MgO, we identify the $(1\bar{1}0)[110]$ slip direction as a key facilitator of plastic deformation, enabled by the presence of alternating charged atomic layers. This simple structural motif supports a bonding relay mechanism that allows for sequential bond rupture and reformation, while preserving structural integrity during dislocation glide. Our results provide direct evidence that oxide plasticity is not only possible but also tunable through crystallographic design. These insights offer a unifying framework for understanding deformation mechanisms in oxides and open new avenues for engineering mechanically resilient oxide-based materials for flexible electronic and structural applications.



\newpage
\clearpage
\noindent \textbf{Methods}\\
\textbf{First-principles calculations}
All density functional theory (DFT) calculations are carried out using the projector augmented wave (PAW) method implemented in the VASP code~\cite{Kresse96p11169}. The exchange-correlation functional is treated within the generalized gradient approximation (GGA)~\cite{John96p3865}. The plane-wave energy cutoff is set to 600~eV. Monkhorst-Pack $k$-point grids~\cite{Monkhorst76p5188} of $2 \times 3 \times 1$, $3 \times 2 \times 1$, and $3 \times 2 \times 1$ are used to sample the Brillouin zones of the $(1\bar{1}0)[110]$ (80 atoms), $(1\bar{1}0)[001]$ (80 atoms), and (001)[100] (75 atoms) SrTiO$_3$ supercells, respectively, while a $6 \times 6 \times 6$ grid is employed for the bulk unit cell. The convergence criteria for energy and force are set to $1 \times 10^{-6}$~eV and $1.0 \times 10^{-3}$~eV/\AA, respectively.
Crystal Orbital Hamilton Population (COHP) and Integrated COHP (ICOHP) analyses are performed using the LOBSTER code~\cite{Richard93p8617, Volker11p5461, Stefan16p1030}. For shear strain-stress curve calculations, at each strain step, shear stress is evaluated by relaxing both the atomic positions and the supercell shape through constrained structural optimization. The shear strain direction is fixed, while all other strain components and atomic coordinates are relaxed until the forces on all atoms fall below $1.0 \times 10^{-2}$~eV/\AA{} and residual stresses (excluding the shear direction) are below 20~MPa. To ensure a quasi-static deformation path, each step begins from the relaxed configuration of the previous one.
Slip energy barriers are determined by fully relaxing all internal atomic positions, except for the outermost layers, which are constrained to maintain the imposed sliding displacement. 

\noindent \textbf{Molecular dynamics simulations}
Molecular dynamics (MD) simulations are performed using the deep potential (DP) model to investigate the deformation behavior of SrTiO$_3$ along the $(1\bar{1}0)[110]$, $(1\bar{1}0)[001]$, and (001)[100] directions. 
The DP model is trained on a database of DFT energies and atomic forces of 39945 configurations of SrTiO$_3$. 
All simulations are carried out using the LAMMPS code~\cite{Plimpton95p1}. The temperature is maintained at 300~K, with a time step of 1~fs, and a constant strain rate of $10^{10}$~s$^{-1}$ is applied during deformation.
The supercells used for these orientations contain 64000, 44640, and 36000 atoms, respectively.
For simulations involving pre-embedded single dislocation glide, larger supercells are constructed, containing 244320, 85695, and 122511 atoms for the $(1\bar{1}0)[110]$, $(1\bar{1}0)[001]$, and (001)[100] directions, respectively. A constant shear velocity of 10~m/s is applied in these dislocation glide simulations. Additional computational details for MgO are provided in the Supplementary Materials.

\newpage
\clearpage
\begin{figure}[t]
\includegraphics[width=1.0\textwidth]{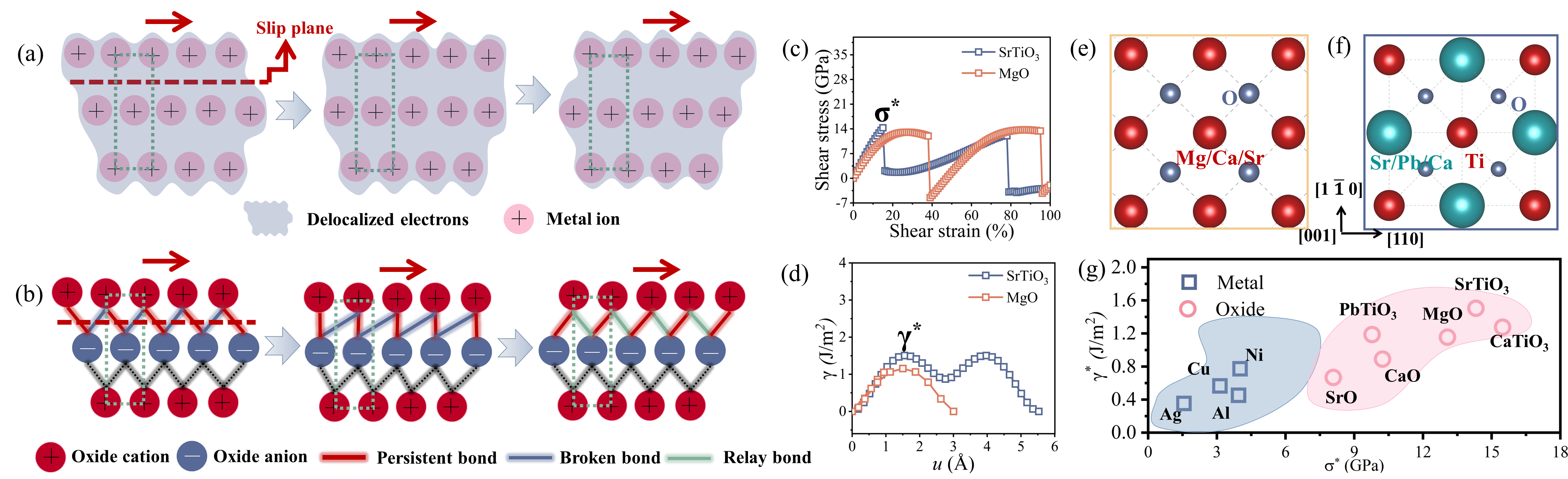}
\caption{
{\bf Oxide plasticity via the bonding relay mechanism.}  
(a) Schematic illustration of lattice slip in metals, where delocalized electrons allows for easy atomic glide.
(b) Schematic of lattice slip in oxides with specific crystallographic orientations that enable the bonding relay mechanism.  
(c) Shear stress–strain curves and (d) slip energy barriers for SrTiO$_3$ and MgO along the (1$\Bar{1}$0)[110] orientation, obtained from DFT calculations.  
(e) Atomic structure of binary oxides (MgO, CaO, and SrO) and (f) ternary perovskite oxides (SrTiO$_3$, CaTiO$_3$, and PbTiO$_3$) along the (1$\Bar{1}$0)[110] slip direction, highlighting alternating charged layers.  
(g) Comparison of maximum shear stress ($\sigma^*$) and maximum slip energy barrier ($\gamma^*$) for selected oxides (1$\bar{1}$0)[110] crystallographic orientation (SrTiO$_3$, PbTiO$_3$, CaTiO$_3$, MgO, CaO, and SrO) and face-centered cubic metals along the (111)[1$\bar{1}$0] crystallographic orientation (Al, Cu, Ni, and Ag).
} 
\label{fig:all}
\end{figure}

\begin{figure}[t]
\includegraphics[width=1.0\textwidth]{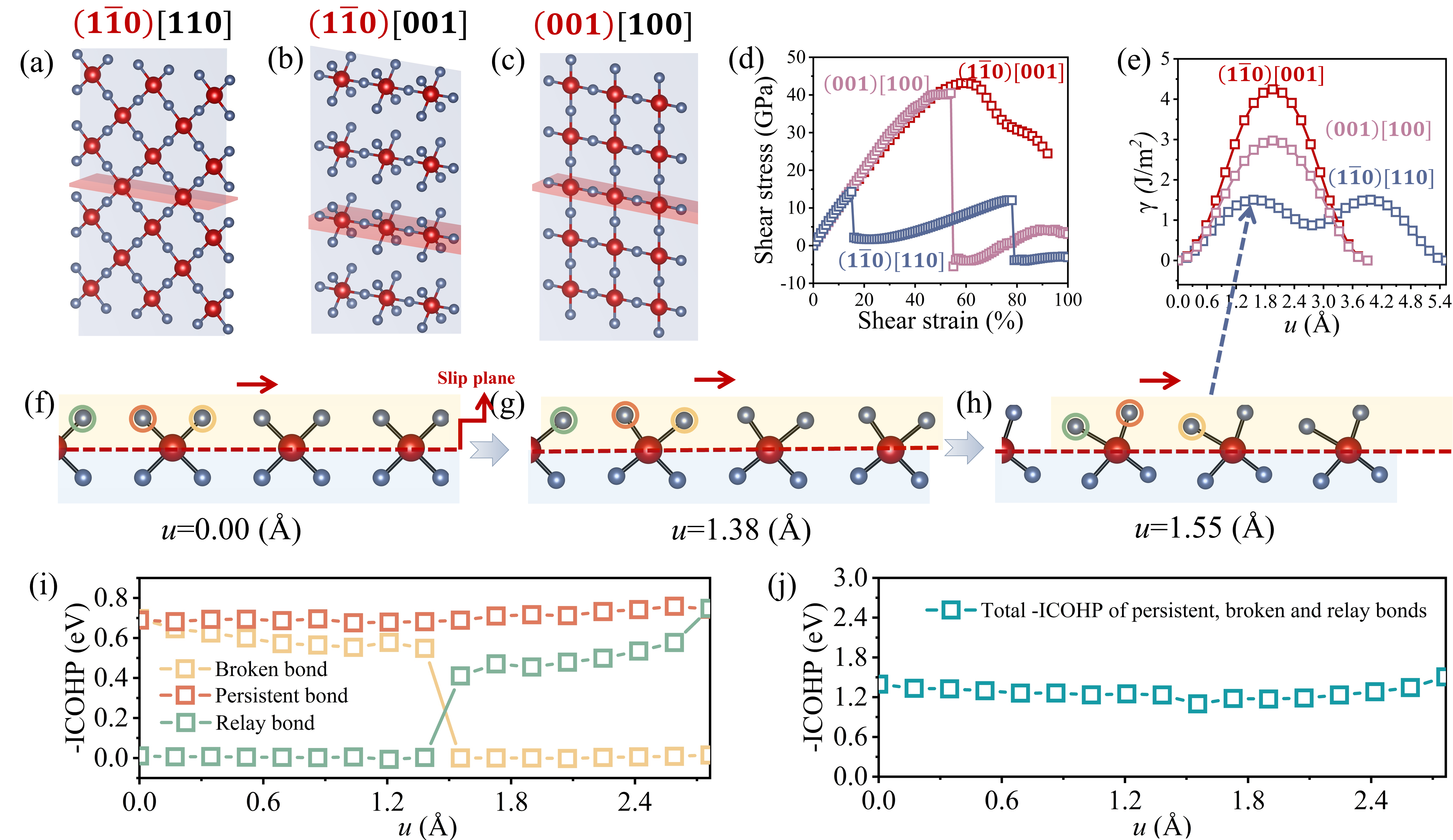}
\caption{
{\bf Plastic deformation via dislocation motion in ($1\bar{1}0$)[110] SrTiO$_3$.}  
Atomic geometry of the Ti--O sublattice in SrTiO$_3$ for three crystallographic orientations: (a) ($1\bar{1}0$)[110], (b) ($1\bar{1}0$)[001], and (c) (001)[100]. Large and small spheres represent Ti and O atoms, respectively.  
(d) Shear stress–strain curves and (e) slip energy barriers obtained from DFT calculations for each orientation.  
(f-h) Snapshots of atomic configurations during shear deformation along the [110] direction for the ($1\bar{1}0$) slip plane at increasing displacement values \textit{u}.
(i) Evolution of --ICOHP values for persist, broken, and relay Ti-O bonds during sliding. Oxygen atoms involved in key Ti--O interactions are highlighted in (f-h). 
(j) Sum of --ICHOP value for all three bond types, showing the overall conservation of bond strength throughout the bonding relay process.
} 
\label{fig:all}
\end{figure}

\begin{figure}[t]
\includegraphics[width=1.0\textwidth]{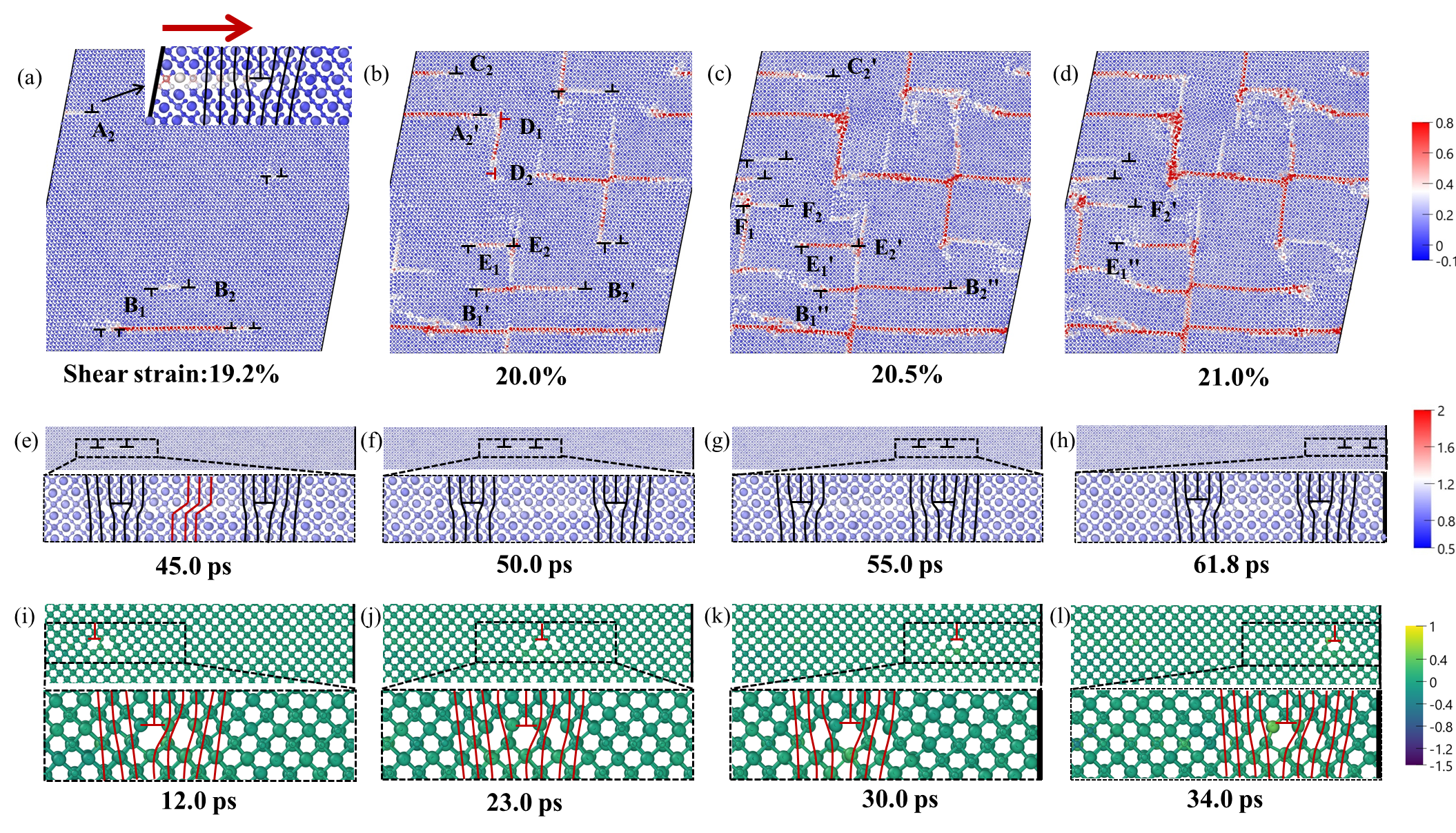}
\caption{
\textbf{Dislocation motion mechanism in $(1\bar{1}0)[110]$ SrTiO$_3$ and MgO.}  
(a–d) Microstructural evolution of bulk SrTiO$_3$ under $(1\bar{1}0)[110]$ shear deformation from MD simulations.  
Time-resolved snapshots of a single pre-introduced dislocation in (e–h) SrTiO$_3$ and (i–l) MgO during shear deformation along the $(1\bar{1}0)[110]$ slip direction. The burgers vector of the paired partial dislocations in (a) to (d) are $\vec{b} = -\frac{a}{2}[110]$ (Burgers vector points to the left) and $\vec{b} = \frac{a}{2}[110]$ (Burgers vector points to the right), respectively. The burgers vector of the paired partial dislocations in (e) to (h) are $\vec{b} = \frac{a}{2}[110]$. The anti-phase boundary (APB) in dislocation cores are presented by a red line in (e). Atomic sizes correspond to elements: Sr (largest), Ti (medium), O (small) in SrTiO$_3$; Mg (largest) and O (small) in MgO. The color bar in (a) to (d) presents the magnitude of shear strain, while the color bar in (e) to (l) denotes the magnitude of strain tensor.
} 
\label{fig:all}
\end{figure}

\begin{figure}[t]
\includegraphics[width=1.0\textwidth]{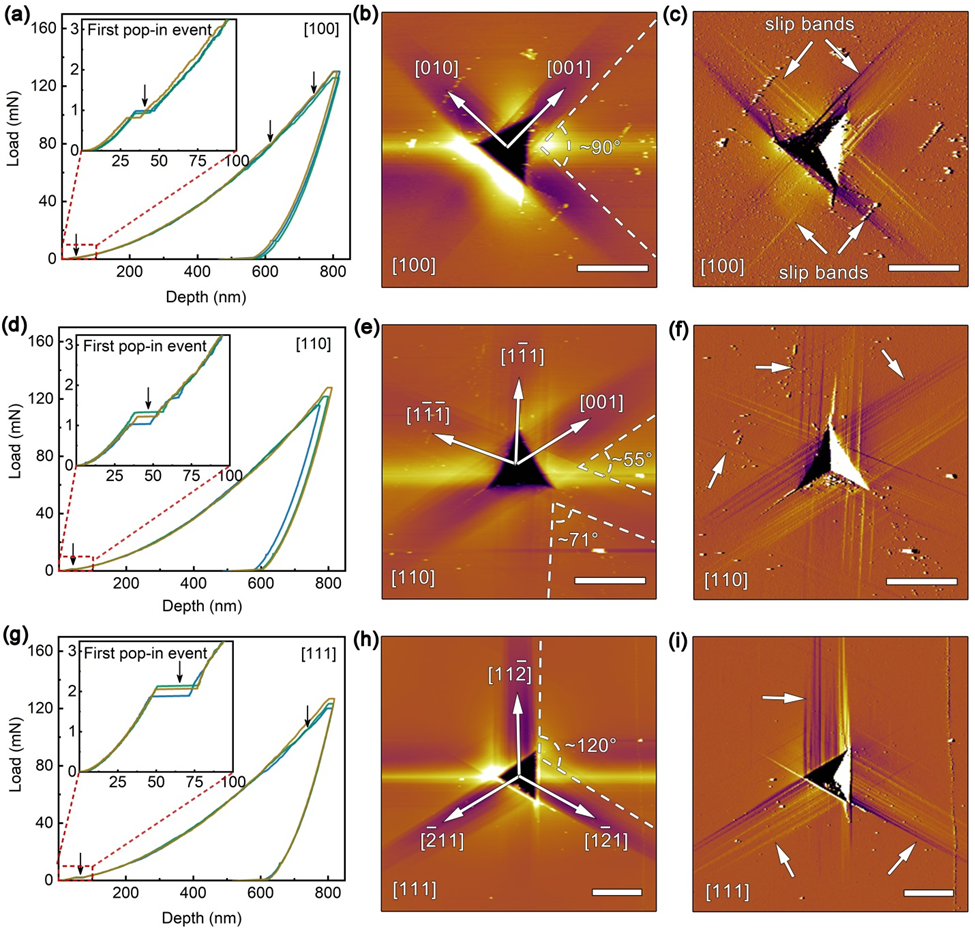}
\caption{
(a) Load-depth curves of nanoindentation on [100]-oriented SrTiO$_3$ single-crystals. Multiple pop-in events were indicated by black arrows. The embedded panel shows the magnified first pop-in event, indicating the occurring of plastic deformation related to the a$\langle110\rangle$ type dislocation motion. (b) AFM surface topography image of an indentation on the [100]-oriented SrTiO$_3$ single-crystal using a Berkovich indenter. (c) Out-of-plane PFM amplitude image corresponding to (b). (100) surface displays clear slip bands parallel to [001] and [010] directions. (d)-(f) and (g)-(i), Load-depth curves of nanoindentation, AFM and PFM images of the [110] and [111]-oriented SrTiO$_3$ single-crystals. Scale bar: 5 $\mu$m.
} 
\label{fig:all}
\end{figure}

\clearpage
\bibliography{SL.bib}

\begin{thebibliography}{25}%
\makeatletter
\providecommand \@ifxundefined [1]{%
 \@ifx{#1\undefined}
}%
\providecommand \@ifnum [1]{%
 \ifnum #1\expandafter \@firstoftwo
 \else \expandafter \@secondoftwo
 \fi
}%
\providecommand \@ifx [1]{%
 \ifx #1\expandafter \@firstoftwo
 \else \expandafter \@secondoftwo
 \fi
}%
\providecommand \natexlab [1]{#1}%
\providecommand \enquote  [1]{``#1''}%
\providecommand \bibnamefont  [1]{#1}%
\providecommand \bibfnamefont [1]{#1}%
\providecommand \citenamefont [1]{#1}%
\providecommand \href@noop [0]{\@secondoftwo}%
\providecommand \href [0]{\begingroup \@sanitize@url \@href}%
\providecommand \@href[1]{\@@startlink{#1}\@@href}%
\providecommand \@@href[1]{\endgroup#1\@@endlink}%
\providecommand \@sanitize@url [0]{\catcode `\\12\catcode `\$12\catcode `\&12\catcode `\#12\catcode `\^12\catcode `\_12\catcode `\%12\relax}%
\providecommand \@@startlink[1]{}%
\providecommand \@@endlink[0]{}%
\providecommand \url  [0]{\begingroup\@sanitize@url \@url }%
\providecommand \@url [1]{\endgroup\@href {#1}{\urlprefix }}%
\providecommand \urlprefix  [0]{URL }%
\providecommand \Eprint [0]{\href }%
\providecommand \doibase [0]{https://doi.org/}%
\providecommand \selectlanguage [0]{\@gobble}%
\providecommand \bibinfo  [0]{\@secondoftwo}%
\providecommand \bibfield  [0]{\@secondoftwo}%
\providecommand \translation [1]{[#1]}%
\providecommand \BibitemOpen [0]{}%
\providecommand \bibitemStop [0]{}%
\providecommand \bibitemNoStop [0]{.\EOS\space}%
\providecommand \EOS [0]{\spacefactor3000\relax}%
\providecommand \BibitemShut  [1]{\csname bibitem#1\endcsname}%
\let\auto@bib@innerbib\@empty
\bibitem [{\citenamefont {Barrett}\ and\ \citenamefont {Massalski}(1992)}]{C.S.Barrett92}%
  \BibitemOpen
  \bibfield  {author} {\bibinfo {author} {\bibfnamefont {C.~S.}\ \bibnamefont {Barrett}}\ and\ \bibinfo {author} {\bibfnamefont {T.~B.}\ \bibnamefont {Massalski}},\ }\href@noop {} {\emph {\bibinfo {title} {Structure of Metals, Third Edition: Crystallographic Methods, Principles and Data}}}\ (\bibinfo  {publisher} {Pergamon, Oxford, {UK}},\ \bibinfo {year} {1992})\BibitemShut {NoStop}%
\bibitem [{\citenamefont {Smithells}\ and\ \citenamefont {J}(1992)}]{Smithells92}%
  \BibitemOpen
  \bibfield  {author} {\bibinfo {author} {\bibnamefont {Smithells}}\ and\ \bibinfo {author} {\bibfnamefont {C.}~\bibnamefont {J}},\ }\href@noop {} {\emph {\bibinfo {title} {Smithells Metals Reference Book (Seventh Edition)}}}\ (\bibinfo  {publisher} {Butterworth-Heinemann, Boston, MA, USA},\ \bibinfo {year} {1992})\BibitemShut {NoStop}%
\bibitem [{\citenamefont {Green}(1998)}]{Green98}%
  \BibitemOpen
  \bibfield  {author} {\bibinfo {author} {\bibfnamefont {D.~J.}\ \bibnamefont {Green}},\ }\href@noop {} {\emph {\bibinfo {title} {An Introduction to the Mechanical Properties of Ceramics}}}\ (\bibinfo  {publisher} {Cambridge University Press, Cambridge, UK, 1998},\ \bibinfo {year} {1998})\BibitemShut {NoStop}%
\bibitem [{\citenamefont {Yi}\ \emph {et~al.}(2022)\citenamefont {Yi}, \citenamefont {Xiangyang}, \citenamefont {Peng}, \citenamefont {Yi}, \citenamefont {Muzhang},\ and\ \citenamefont {Chunlei}}]{Li22p6925}%
  \BibitemOpen
  \bibfield  {author} {\bibinfo {author} {\bibfnamefont {L.}~\bibnamefont {Yi}}, \bibinfo {author} {\bibfnamefont {L.}~\bibnamefont {Xiangyang}}, \bibinfo {author} {\bibfnamefont {Z.}~\bibnamefont {Peng}}, \bibinfo {author} {\bibfnamefont {H.}~\bibnamefont {Yi}}, \bibinfo {author} {\bibfnamefont {H.}~\bibnamefont {Muzhang}},\ and\ \bibinfo {author} {\bibfnamefont {W.}~\bibnamefont {Chunlei}},\ }\bibfield  {title} {\bibinfo {title} {Theoretical insights into the peierls plasticity in {SrTiO}$_3$ ceramics via dislocation remodelling},\ }\href {https://doi.org/10.1038/s41467-022-34741-4} {\bibfield  {journal} {\bibinfo  {journal} {Nat. Commun.}\ }\textbf {\bibinfo {volume} {13}},\ \bibinfo {pages} {6925} (\bibinfo {year} {2022})}\BibitemShut {NoStop}%
\bibitem [{\citenamefont {Lukas}\ \emph {et~al.}(2021)\citenamefont {Lukas}, \citenamefont {J}, \citenamefont {Xufei}, \citenamefont {Ning}, \citenamefont {Can}, \citenamefont {Carsten}, \citenamefont {Enrico}, \citenamefont {Marion}, \citenamefont {Wolfgang}, \citenamefont {A}, \citenamefont {Peng}, \citenamefont {Karsten}, \citenamefont {Atsutomo}, \citenamefont {Karsten}, \citenamefont {Hugh},\ and\ \citenamefont {Jürgen}}]{Porz21p1528}%
  \BibitemOpen
  \bibfield  {author} {\bibinfo {author} {\bibfnamefont {P.}~\bibnamefont {Lukas}}, \bibinfo {author} {\bibfnamefont {K.~A.}\ \bibnamefont {J}}, \bibinfo {author} {\bibfnamefont {F.}~\bibnamefont {Xufei}}, \bibinfo {author} {\bibfnamefont {L.}~\bibnamefont {Ning}}, \bibinfo {author} {\bibfnamefont {Y.}~\bibnamefont {Can}}, \bibinfo {author} {\bibfnamefont {D.}~\bibnamefont {Carsten}}, \bibinfo {author} {\bibfnamefont {B.}~\bibnamefont {Enrico}}, \bibinfo {author} {\bibfnamefont {H.}~\bibnamefont {Marion}}, \bibinfo {author} {\bibfnamefont {R.}~\bibnamefont {Wolfgang}}, \bibinfo {author} {\bibfnamefont {P.~E.}\ \bibnamefont {A}}, \bibinfo {author} {\bibfnamefont {G.}~\bibnamefont {Peng}}, \bibinfo {author} {\bibfnamefont {D.}~\bibnamefont {Karsten}}, \bibinfo {author} {\bibfnamefont {N.}~\bibnamefont {Atsutomo}}, \bibinfo {author} {\bibfnamefont {A.}~\bibnamefont {Karsten}}, \bibinfo {author} {\bibfnamefont {S.}~\bibnamefont {Hugh}},\ and\ \bibinfo {author} {\bibfnamefont {R.}~\bibnamefont {Jürgen}},\
  }\bibfield  {title} {\bibinfo {title} {Dislocation-toughened ceramics},\ }\href {https://doi.org/10.1038/s41467-024-51615-z} {\bibfield  {journal} {\bibinfo  {journal} {Mater. Horiz.}\ }\textbf {\bibinfo {volume} {8}},\ \bibinfo {pages} {1528} (\bibinfo {year} {2021})}\BibitemShut {NoStop}%
\bibitem [{\citenamefont {Jin}\ \emph {et~al.}(2019)\citenamefont {Jin}, \citenamefont {Jaehun}, \citenamefont {Jie}, \citenamefont {Harry}, \citenamefont {Sichuang}, \citenamefont {Han}, \citenamefont {Li}, \citenamefont {Jie}, \citenamefont {Xuejing}, \citenamefont {Colin}, \citenamefont {Thomas}, \citenamefont {Edwin}, \citenamefont {K}, \citenamefont {Noam}, \citenamefont {Stephen}, \citenamefont {Haiyan},\ and\ \citenamefont {Xinghang}}]{Li19p5519}%
  \BibitemOpen
  \bibfield  {author} {\bibinfo {author} {\bibfnamefont {L.}~\bibnamefont {Jin}}, \bibinfo {author} {\bibfnamefont {C.}~\bibnamefont {Jaehun}}, \bibinfo {author} {\bibfnamefont {D.}~\bibnamefont {Jie}}, \bibinfo {author} {\bibfnamefont {C.}~\bibnamefont {Harry}}, \bibinfo {author} {\bibfnamefont {X.}~\bibnamefont {Sichuang}}, \bibinfo {author} {\bibfnamefont {W.}~\bibnamefont {Han}}, \bibinfo {author} {\bibfnamefont {P.~X.}\ \bibnamefont {Li}}, \bibinfo {author} {\bibfnamefont {J.}~\bibnamefont {Jie}}, \bibinfo {author} {\bibfnamefont {W.}~\bibnamefont {Xuejing}}, \bibinfo {author} {\bibfnamefont {O.}~\bibnamefont {Colin}}, \bibinfo {author} {\bibfnamefont {T.}~\bibnamefont {Thomas}}, \bibinfo {author} {\bibfnamefont {G.~R.}\ \bibnamefont {Edwin}}, \bibinfo {author} {\bibfnamefont {M.~A.}\ \bibnamefont {K}}, \bibinfo {author} {\bibfnamefont {B.}~\bibnamefont {Noam}}, \bibinfo {author} {\bibfnamefont {H.~C.}\ \bibnamefont {Stephen}}, \bibinfo {author} {\bibfnamefont {W.}~\bibnamefont {Haiyan}},\ and\ \bibinfo
  {author} {\bibfnamefont {Z.}~\bibnamefont {Xinghang}},\ }\bibfield  {title} {\bibinfo {title} {Nanoscale stacking fault–assisted room temperature plasticity in flash-sintered {TiO}$_{\textrm{2}}$},\ }\href {https://doi.org/10.1126/sciadv.aaw5519} {\bibfield  {journal} {\bibinfo  {journal} {Sci. Adv.}\ }\textbf {\bibinfo {volume} {5}},\ \bibinfo {pages} {5519} (\bibinfo {year} {2019})}\BibitemShut {NoStop}%
\bibitem [{\citenamefont {Trung}\ and\ \citenamefont {Lee}(2016)}]{Trung16p4338}%
  \BibitemOpen
  \bibfield  {author} {\bibinfo {author} {\bibfnamefont {T.}~\bibnamefont {Trung}}\ and\ \bibinfo {author} {\bibfnamefont {N.-E.}\ \bibnamefont {Lee}},\ }\bibfield  {title} {\bibinfo {title} {Flexible and stretchable physical sensor integrated platforms for wearable human-activity monitoringand personal healthcare},\ }\href {https://doi.org/10.1002/adma.201504244} {\bibfield  {journal} {\bibinfo  {journal} {Adv. Mater.}\ }\textbf {\bibinfo {volume} {28}},\ \bibinfo {pages} {4338} (\bibinfo {year} {2016})}\BibitemShut {NoStop}%
\bibitem [{\citenamefont {Schroeder}\ \emph {et~al.}(2022)\citenamefont {Schroeder}, \citenamefont {Park}, \citenamefont {Mikolajick},\ and\ \citenamefont {Hwang}}]{Schroeder22p653}%
  \BibitemOpen
  \bibfield  {author} {\bibinfo {author} {\bibfnamefont {U.}~\bibnamefont {Schroeder}}, \bibinfo {author} {\bibfnamefont {M.~H.}\ \bibnamefont {Park}}, \bibinfo {author} {\bibfnamefont {T.}~\bibnamefont {Mikolajick}},\ and\ \bibinfo {author} {\bibfnamefont {C.~S.}\ \bibnamefont {Hwang}},\ }\bibfield  {title} {\bibinfo {title} {The fundamentals and applications of ferroelectric {HfO$_2$}},\ }\href {https://doi.org/10.1038/s41578-022-00431-2} {\bibfield  {journal} {\bibinfo  {journal} {Nat. Rev. Mater.}\ }\textbf {\bibinfo {volume} {7}},\ \bibinfo {pages} {653–669} (\bibinfo {year} {2022})}\BibitemShut {NoStop}%
\bibitem [{\citenamefont {Coll}\ \emph {et~al.}(2019)\citenamefont {Coll}, \citenamefont {Fontcuberta}, \citenamefont {Althammer}, \citenamefont {Bibes}, \citenamefont {Boschker}, \citenamefont {Calleja}, \citenamefont {Cheng}, \citenamefont {Cuoco}, \citenamefont {Dittmann}, \citenamefont {Dkhil}, \citenamefont {El~Baggari}, \citenamefont {Fanciulli}, \citenamefont {Fina}, \citenamefont {Fortunato}, \citenamefont {Frontera}, \citenamefont {Fujita}, \citenamefont {Garcia}, \citenamefont {Goennenwein}, \citenamefont {Granqvist}, \citenamefont {Grollier}, \citenamefont {Gross}, \citenamefont {Hagfeldt}, \citenamefont {Herranz}, \citenamefont {Hono}, \citenamefont {Houwman}, \citenamefont {Huijben}, \citenamefont {Kalaboukhov}, \citenamefont {Keeble}, \citenamefont {Koster}, \citenamefont {Kourkoutis}, \citenamefont {Levy}, \citenamefont {Lira-Cantu}, \citenamefont {MacManus-Driscoll}, \citenamefont {Mannhart}, \citenamefont {Martins}, \citenamefont {Menzel}, \citenamefont {Mikolajick}, \citenamefont {Napari},
  \citenamefont {Nguyen}, \citenamefont {Niklasson}, \citenamefont {Paillard}, \citenamefont {Panigrahi}, \citenamefont {Rijnders}, \citenamefont {Sánchez}, \citenamefont {Sanchis}, \citenamefont {Sanna}, \citenamefont {Schlom}, \citenamefont {Schroeder}, \citenamefont {Shen}, \citenamefont {Siemon}, \citenamefont {Spreitzer}, \citenamefont {Sukegawa}, \citenamefont {Tamayo}, \citenamefont {van~den Brink}, \citenamefont {Pryds},\ and\ \citenamefont {Granozio}}]{Coll19p1}%
  \BibitemOpen
  \bibfield  {author} {\bibinfo {author} {\bibfnamefont {M.}~\bibnamefont {Coll}}, \bibinfo {author} {\bibfnamefont {J.}~\bibnamefont {Fontcuberta}}, \bibinfo {author} {\bibfnamefont {M.}~\bibnamefont {Althammer}}, \bibinfo {author} {\bibfnamefont {M.}~\bibnamefont {Bibes}}, \bibinfo {author} {\bibfnamefont {H.}~\bibnamefont {Boschker}}, \bibinfo {author} {\bibfnamefont {A.}~\bibnamefont {Calleja}}, \bibinfo {author} {\bibfnamefont {G.}~\bibnamefont {Cheng}}, \bibinfo {author} {\bibfnamefont {M.}~\bibnamefont {Cuoco}}, \bibinfo {author} {\bibfnamefont {R.}~\bibnamefont {Dittmann}}, \bibinfo {author} {\bibfnamefont {B.}~\bibnamefont {Dkhil}}, \bibinfo {author} {\bibfnamefont {I.}~\bibnamefont {El~Baggari}}, \bibinfo {author} {\bibfnamefont {M.}~\bibnamefont {Fanciulli}}, \bibinfo {author} {\bibfnamefont {I.}~\bibnamefont {Fina}}, \bibinfo {author} {\bibfnamefont {E.}~\bibnamefont {Fortunato}}, \bibinfo {author} {\bibfnamefont {C.}~\bibnamefont {Frontera}}, \bibinfo {author} {\bibfnamefont {S.}~\bibnamefont
  {Fujita}}, \bibinfo {author} {\bibfnamefont {V.}~\bibnamefont {Garcia}}, \bibinfo {author} {\bibfnamefont {S.}~\bibnamefont {Goennenwein}}, \bibinfo {author} {\bibfnamefont {C.-G.}\ \bibnamefont {Granqvist}}, \bibinfo {author} {\bibfnamefont {J.}~\bibnamefont {Grollier}}, \bibinfo {author} {\bibfnamefont {R.}~\bibnamefont {Gross}}, \bibinfo {author} {\bibfnamefont {A.}~\bibnamefont {Hagfeldt}}, \bibinfo {author} {\bibfnamefont {G.}~\bibnamefont {Herranz}}, \bibinfo {author} {\bibfnamefont {K.}~\bibnamefont {Hono}}, \bibinfo {author} {\bibfnamefont {E.}~\bibnamefont {Houwman}}, \bibinfo {author} {\bibfnamefont {M.}~\bibnamefont {Huijben}}, \bibinfo {author} {\bibfnamefont {A.}~\bibnamefont {Kalaboukhov}}, \bibinfo {author} {\bibfnamefont {D.}~\bibnamefont {Keeble}}, \bibinfo {author} {\bibfnamefont {G.}~\bibnamefont {Koster}}, \bibinfo {author} {\bibfnamefont {L.}~\bibnamefont {Kourkoutis}}, \bibinfo {author} {\bibfnamefont {J.}~\bibnamefont {Levy}}, \bibinfo {author} {\bibfnamefont {M.}~\bibnamefont
  {Lira-Cantu}}, \bibinfo {author} {\bibfnamefont {J.}~\bibnamefont {MacManus-Driscoll}}, \bibinfo {author} {\bibfnamefont {J.}~\bibnamefont {Mannhart}}, \bibinfo {author} {\bibfnamefont {R.}~\bibnamefont {Martins}}, \bibinfo {author} {\bibfnamefont {S.}~\bibnamefont {Menzel}}, \bibinfo {author} {\bibfnamefont {T.}~\bibnamefont {Mikolajick}}, \bibinfo {author} {\bibfnamefont {M.}~\bibnamefont {Napari}}, \bibinfo {author} {\bibfnamefont {M.}~\bibnamefont {Nguyen}}, \bibinfo {author} {\bibfnamefont {G.}~\bibnamefont {Niklasson}}, \bibinfo {author} {\bibfnamefont {C.}~\bibnamefont {Paillard}}, \bibinfo {author} {\bibfnamefont {S.}~\bibnamefont {Panigrahi}}, \bibinfo {author} {\bibfnamefont {G.}~\bibnamefont {Rijnders}}, \bibinfo {author} {\bibfnamefont {F.}~\bibnamefont {Sánchez}}, \bibinfo {author} {\bibfnamefont {P.}~\bibnamefont {Sanchis}}, \bibinfo {author} {\bibfnamefont {S.}~\bibnamefont {Sanna}}, \bibinfo {author} {\bibfnamefont {D.}~\bibnamefont {Schlom}}, \bibinfo {author} {\bibfnamefont
  {U.}~\bibnamefont {Schroeder}}, \bibinfo {author} {\bibfnamefont {K.}~\bibnamefont {Shen}}, \bibinfo {author} {\bibfnamefont {A.}~\bibnamefont {Siemon}}, \bibinfo {author} {\bibfnamefont {M.}~\bibnamefont {Spreitzer}}, \bibinfo {author} {\bibfnamefont {H.}~\bibnamefont {Sukegawa}}, \bibinfo {author} {\bibfnamefont {R.}~\bibnamefont {Tamayo}}, \bibinfo {author} {\bibfnamefont {J.}~\bibnamefont {van~den Brink}}, \bibinfo {author} {\bibfnamefont {N.}~\bibnamefont {Pryds}},\ and\ \bibinfo {author} {\bibfnamefont {F.~M.}\ \bibnamefont {Granozio}},\ }\bibfield  {title} {\bibinfo {title} {Towards oxide electronics: a roadmap},\ }\href {https://doi.org/10.1016/j.apsusc.2019.03.312} {\bibfield  {journal} {\bibinfo  {journal} {Appl. Surf. Sci.}\ }\textbf {\bibinfo {volume} {482}},\ \bibinfo {pages} {1–93} (\bibinfo {year} {2019})}\BibitemShut {NoStop}%
\bibitem [{\citenamefont {Xun}\ \emph {et~al.}(2018)\citenamefont {Xun}, \citenamefont {Hongyi}, \citenamefont {Feng}, \citenamefont {Ruiheng}, \citenamefont {Tuo}, \citenamefont {Pengfei}, \citenamefont {Ulrich}, \citenamefont {Yuri},\ and\ \citenamefont {Lidong}}]{Shi18p421}%
  \BibitemOpen
  \bibfield  {author} {\bibinfo {author} {\bibfnamefont {S.}~\bibnamefont {Xun}}, \bibinfo {author} {\bibfnamefont {C.}~\bibnamefont {Hongyi}}, \bibinfo {author} {\bibfnamefont {H.}~\bibnamefont {Feng}}, \bibinfo {author} {\bibfnamefont {L.}~\bibnamefont {Ruiheng}}, \bibinfo {author} {\bibfnamefont {W.}~\bibnamefont {Tuo}}, \bibinfo {author} {\bibfnamefont {Q.}~\bibnamefont {Pengfei}}, \bibinfo {author} {\bibfnamefont {B.}~\bibnamefont {Ulrich}}, \bibinfo {author} {\bibfnamefont {G.}~\bibnamefont {Yuri}},\ and\ \bibinfo {author} {\bibfnamefont {C.}~\bibnamefont {Lidong}},\ }\bibfield  {title} {\bibinfo {title} {Room-temperature ductile inorganic semiconductor},\ }\href {https://doi.org/10.1038/s41563-018-0047-z} {\bibfield  {journal} {\bibinfo  {journal} {Nature. Mater.}\ }\textbf {\bibinfo {volume} {17}},\ \bibinfo {pages} {421} (\bibinfo {year} {2018})}\BibitemShut {NoStop}%
\bibitem [{\citenamefont {Yu}\ \emph {et~al.}(2018)\citenamefont {Yu}, \citenamefont {Atsutomo},\ and\ \citenamefont {Katsuyuki}}]{Oshima18p772}%
  \BibitemOpen
  \bibfield  {author} {\bibinfo {author} {\bibfnamefont {O.}~\bibnamefont {Yu}}, \bibinfo {author} {\bibfnamefont {N.}~\bibnamefont {Atsutomo}},\ and\ \bibinfo {author} {\bibfnamefont {M.}~\bibnamefont {Katsuyuki}},\ }\bibfield  {title} {\bibinfo {title} {Extraordinary plasticity of an inorganic semiconductor in darkness},\ }\href {https://doi.org/10.1126/science.aar6035} {\bibfield  {journal} {\bibinfo  {journal} {Science}\ }\textbf {\bibinfo {volume} {360}},\ \bibinfo {pages} {772} (\bibinfo {year} {2018})}\BibitemShut {NoStop}%
\bibitem [{\citenamefont {Peng}\ \emph {et~al.}(2024)\citenamefont {Peng}, \citenamefont {Wenhua}, \citenamefont {Yue}, \citenamefont {Shizhen}, \citenamefont {Xiaojing}, \citenamefont {Jiamin}, \citenamefont {Tianyu}, \citenamefont {Sheng}, \citenamefont {Huimin}, \citenamefont {Jinxuan}, \citenamefont {Xiaodong}, \citenamefont {Shuaihang}, \citenamefont {Lijia}, \citenamefont {Guoqiang}, \citenamefont {Feng}, \citenamefont {Xingjun}, \citenamefont {Jun}, \citenamefont {Yuhao}, \citenamefont {Yumei},\ and\ \citenamefont {Qian}}]{Zhao24p777}%
  \BibitemOpen
  \bibfield  {author} {\bibinfo {author} {\bibfnamefont {Z.}~\bibnamefont {Peng}}, \bibinfo {author} {\bibfnamefont {X.}~\bibnamefont {Wenhua}}, \bibinfo {author} {\bibfnamefont {Z.}~\bibnamefont {Yue}}, \bibinfo {author} {\bibfnamefont {Z.}~\bibnamefont {Shizhen}}, \bibinfo {author} {\bibfnamefont {M.}~\bibnamefont {Xiaojing}}, \bibinfo {author} {\bibfnamefont {Q.}~\bibnamefont {Jiamin}}, \bibinfo {author} {\bibfnamefont {Z.}~\bibnamefont {Tianyu}}, \bibinfo {author} {\bibfnamefont {Y.}~\bibnamefont {Sheng}}, \bibinfo {author} {\bibfnamefont {M.}~\bibnamefont {Huimin}}, \bibinfo {author} {\bibfnamefont {C.}~\bibnamefont {Jinxuan}}, \bibinfo {author} {\bibfnamefont {W.}~\bibnamefont {Xiaodong}}, \bibinfo {author} {\bibfnamefont {H.}~\bibnamefont {Shuaihang}}, \bibinfo {author} {\bibfnamefont {Z.}~\bibnamefont {Lijia}}, \bibinfo {author} {\bibfnamefont {X.}~\bibnamefont {Guoqiang}}, \bibinfo {author} {\bibfnamefont {C.}~\bibnamefont {Feng}}, \bibinfo {author} {\bibfnamefont {L.}~\bibnamefont {Xingjun}},
  \bibinfo {author} {\bibfnamefont {M.}~\bibnamefont {Jun}}, \bibinfo {author} {\bibfnamefont {F.}~\bibnamefont {Yuhao}}, \bibinfo {author} {\bibfnamefont {W.}~\bibnamefont {Yumei}},\ and\ \bibinfo {author} {\bibfnamefont {Z.}~\bibnamefont {Qian}},\ }\bibfield  {title} {\bibinfo {title} {Plasticity in single-crystalline {Mg$_3$Bi$_2$} thermoelectric material},\ }\href {https://doi.org/10.1038/s41586-024-07621-8} {\bibfield  {journal} {\bibinfo  {journal} {Nature}\ }\textbf {\bibinfo {volume} {631}},\ \bibinfo {pages} {777} (\bibinfo {year} {2024})}\BibitemShut {NoStop}%
\bibitem [{\citenamefont {Dieter}\ \emph {et~al.}(2001)\citenamefont {Dieter}, \citenamefont {Shahram}, \citenamefont {Wilfried},\ and\ \citenamefont {Manfred}}]{Brunner01p1161}%
  \BibitemOpen
  \bibfield  {author} {\bibinfo {author} {\bibfnamefont {B.}~\bibnamefont {Dieter}}, \bibinfo {author} {\bibfnamefont {T.}~\bibnamefont {Shahram}}, \bibinfo {author} {\bibfnamefont {S.}~\bibnamefont {Wilfried}},\ and\ \bibinfo {author} {\bibfnamefont {R.}~\bibnamefont {Manfred}},\ }\bibfield  {title} {\bibinfo {title} {Surprising results of a study on the plasticity in strontium titanate},\ }\href {https://doi.org/10.1111/j.1151-2916.2001.tb00805.x} {\bibfield  {journal} {\bibinfo  {journal} {J. Am. Ceram. Soc.}\ }\textbf {\bibinfo {volume} {84}},\ \bibinfo {pages} {1161} (\bibinfo {year} {2001})}\BibitemShut {NoStop}%
\bibitem [{\citenamefont {Gumbsch}\ \emph {et~al.}(2001)\citenamefont {Gumbsch}, \citenamefont {Taeri-Baghbadrani}, \citenamefont {Brunner}, \citenamefont {Sigle},\ and\ \citenamefont {R\"uhle}}]{Gumbsch01p085505}%
  \BibitemOpen
  \bibfield  {author} {\bibinfo {author} {\bibfnamefont {P.}~\bibnamefont {Gumbsch}}, \bibinfo {author} {\bibfnamefont {S.}~\bibnamefont {Taeri-Baghbadrani}}, \bibinfo {author} {\bibfnamefont {D.}~\bibnamefont {Brunner}}, \bibinfo {author} {\bibfnamefont {W.}~\bibnamefont {Sigle}},\ and\ \bibinfo {author} {\bibfnamefont {M.}~\bibnamefont {R\"uhle}},\ }\bibfield  {title} {\bibinfo {title} {Plasticity and an inverse brittle-to-ductile transition in strontium titanate},\ }\href {https://doi.org/10.1038/s41467-024-51615-z} {\bibfield  {journal} {\bibinfo  {journal} {Phys. Rev. Lett.}\ }\textbf {\bibinfo {volume} {87}},\ \bibinfo {pages} {085505} (\bibinfo {year} {2001})}\BibitemShut {NoStop}%
\bibitem [{\citenamefont {Kai‐Hsun}\ \emph {et~al.}(2009)\citenamefont {Kai‐Hsun}, \citenamefont {New‐Jin},\ and\ \citenamefont {Hong‐Yang}}]{Yang09p2345}%
  \BibitemOpen
  \bibfield  {author} {\bibinfo {author} {\bibfnamefont {Y.}~\bibnamefont {Kai‐Hsun}}, \bibinfo {author} {\bibfnamefont {H.}~\bibnamefont {New‐Jin}},\ and\ \bibinfo {author} {\bibfnamefont {L.}~\bibnamefont {Hong‐Yang}},\ }\bibfield  {title} {\bibinfo {title} {Deformation microstructure in (001) single crystal strontium titanate by vickers indentation},\ }\href {https://doi.org/10.1111/j.1551-2916.2009.03189.x} {\bibfield  {journal} {\bibinfo  {journal} {J. Am. Ceram. Soc.}\ }\textbf {\bibinfo {volume} {92}},\ \bibinfo {pages} {2345} (\bibinfo {year} {2009})}\BibitemShut {NoStop}%
\bibitem [{\citenamefont {Xufei}\ \emph {et~al.}(2024)\citenamefont {Xufei}, \citenamefont {Wenjun}, \citenamefont {Jiawen}, \citenamefont {Christian}, \citenamefont {Junhua}, \citenamefont {Sebastian}, \citenamefont {Ulrike}, \citenamefont {Atsutomo}, \citenamefont {Karsten},\ and\ \citenamefont {Jürgen}}]{Fang24p81}%
  \BibitemOpen
  \bibfield  {author} {\bibinfo {author} {\bibfnamefont {F.}~\bibnamefont {Xufei}}, \bibinfo {author} {\bibfnamefont {L.}~\bibnamefont {Wenjun}}, \bibinfo {author} {\bibfnamefont {Z.}~\bibnamefont {Jiawen}}, \bibinfo {author} {\bibfnamefont {M.}~\bibnamefont {Christian}}, \bibinfo {author} {\bibfnamefont {H.}~\bibnamefont {Junhua}}, \bibinfo {author} {\bibfnamefont {B.}~\bibnamefont {Sebastian}}, \bibinfo {author} {\bibfnamefont {K.}~\bibnamefont {Ulrike}}, \bibinfo {author} {\bibfnamefont {N.}~\bibnamefont {Atsutomo}}, \bibinfo {author} {\bibfnamefont {D.}~\bibnamefont {Karsten}},\ and\ \bibinfo {author} {\bibfnamefont {R.}~\bibnamefont {Jürgen}},\ }\bibfield  {title} {\bibinfo {title} {Harvesting room-temperature plasticity in ceramics by mechanically seeded dislocations},\ }\href {https://doi.org/10.1016/j.mattod.2024.11.014} {\bibfield  {journal} {\bibinfo  {journal} {Mater. Today.}\ }\textbf {\bibinfo {volume} {82}},\ \bibinfo {pages} {81} (\bibinfo {year} {2024})}\BibitemShut {NoStop}%
\bibitem [{\citenamefont {Xi}\ \emph {et~al.}(2024)\citenamefont {Xi}, \citenamefont {Anirban}, \citenamefont {Bochao}, \citenamefont {Sajna}, \citenamefont {Nadav}, \citenamefont {Shai}, \citenamefont {Luka}, \citenamefont {Liam}, \citenamefont {Alexander}, \citenamefont {Martin}, \citenamefont {Damjan}, \citenamefont {Ilya}, \citenamefont {Beena},\ and\ \citenamefont {Avraham}}]{Wang24p7442}%
  \BibitemOpen
  \bibfield  {author} {\bibinfo {author} {\bibfnamefont {W.}~\bibnamefont {Xi}}, \bibinfo {author} {\bibfnamefont {K.}~\bibnamefont {Anirban}}, \bibinfo {author} {\bibfnamefont {X.}~\bibnamefont {Bochao}}, \bibinfo {author} {\bibfnamefont {H.}~\bibnamefont {Sajna}}, \bibinfo {author} {\bibfnamefont {R.}~\bibnamefont {Nadav}}, \bibinfo {author} {\bibfnamefont {R.}~\bibnamefont {Shai}}, \bibinfo {author} {\bibfnamefont {R.}~\bibnamefont {Luka}}, \bibinfo {author} {\bibfnamefont {T.}~\bibnamefont {Liam}}, \bibinfo {author} {\bibfnamefont {M.}~\bibnamefont {Alexander}}, \bibinfo {author} {\bibfnamefont {G.}~\bibnamefont {Martin}}, \bibinfo {author} {\bibfnamefont {P.}~\bibnamefont {Damjan}}, \bibinfo {author} {\bibfnamefont {S.}~\bibnamefont {Ilya}}, \bibinfo {author} {\bibfnamefont {K.}~\bibnamefont {Beena}},\ and\ \bibinfo {author} {\bibfnamefont {K.}~\bibnamefont {Avraham}},\ }\bibfield  {title} {\bibinfo {title} {Multiferroicity in plastically deformed {SrTiO}$_3$},\ }\href
  {https://doi.org/10.1038/s41467-024-51615-z} {\bibfield  {journal} {\bibinfo  {journal} {Nat. Commun.}\ }\textbf {\bibinfo {volume} {15}},\ \bibinfo {pages} {7442} (\bibinfo {year} {2024})}\BibitemShut {NoStop}%
\bibitem [{\citenamefont {Castillo-Rodríguez}\ and\ \citenamefont {Sigle}(2010)}]{Castillo-Rodríguez10p270}%
  \BibitemOpen
  \bibfield  {author} {\bibinfo {author} {\bibfnamefont {M.}~\bibnamefont {Castillo-Rodríguez}}\ and\ \bibinfo {author} {\bibfnamefont {W.}~\bibnamefont {Sigle}},\ }\bibfield  {title} {\bibinfo {title} {Dislocation dissociation and stacking-fault energy calculation in strontium titanate},\ }\href {https://doi.org/10.1016/j.scriptamat.2009.11.016} {\bibfield  {journal} {\bibinfo  {journal} {Scr. Mater.}\ }\textbf {\bibinfo {volume} {62}},\ \bibinfo {pages} {270} (\bibinfo {year} {2010})}\BibitemShut {NoStop}%
\bibitem [{\citenamefont {Kresse}\ and\ \citenamefont {Furthmüller}(1996)}]{Kresse96p11169}%
  \BibitemOpen
  \bibfield  {author} {\bibinfo {author} {\bibfnamefont {G.}~\bibnamefont {Kresse}}\ and\ \bibinfo {author} {\bibfnamefont {J.}~\bibnamefont {Furthmüller}},\ }\bibfield  {title} {\bibinfo {title} {Efficient iterative schemes for \textit{ab initio} total-energy calculations using a plane-wave basis set},\ }\href {https://doi.org/10.1103/PhysRevB.54.11169} {\bibfield  {journal} {\bibinfo  {journal} {Phys. Rev. B.}\ }\textbf {\bibinfo {volume} {54}},\ \bibinfo {pages} {11169} (\bibinfo {year} {1996})}\BibitemShut {NoStop}%
\bibitem [{\citenamefont {Perdew}\ \emph {et~al.}(1996)\citenamefont {Perdew}, \citenamefont {Burke},\ and\ \citenamefont {Ernzerhof}}]{John96p3865}%
  \BibitemOpen
  \bibfield  {author} {\bibinfo {author} {\bibfnamefont {J.~P.}\ \bibnamefont {Perdew}}, \bibinfo {author} {\bibfnamefont {K.}~\bibnamefont {Burke}},\ and\ \bibinfo {author} {\bibfnamefont {M.}~\bibnamefont {Ernzerhof}},\ }\bibfield  {title} {\bibinfo {title} {Generalized gradient approximation made simple},\ }\href {https://doi.org/10.1103/PhysRevLett.77.3865} {\bibfield  {journal} {\bibinfo  {journal} {Phys. Rev. Lett.}\ }\textbf {\bibinfo {volume} {77}},\ \bibinfo {pages} {3865} (\bibinfo {year} {1996})}\BibitemShut {NoStop}%
\bibitem [{\citenamefont {Monkhorst}\ and\ \citenamefont {Pack}(1976)}]{Monkhorst76p5188}%
  \BibitemOpen
  \bibfield  {author} {\bibinfo {author} {\bibfnamefont {H.~J.}\ \bibnamefont {Monkhorst}}\ and\ \bibinfo {author} {\bibfnamefont {J.~D.}\ \bibnamefont {Pack}},\ }\bibfield  {title} {\bibinfo {title} {Special points for brillouin-zone integrations},\ }\href {https://doi.org/10.1103/PhysRevB.13.5188} {\bibfield  {journal} {\bibinfo  {journal} {Phys. Rev. B.}\ }\textbf {\bibinfo {volume} {13}},\ \bibinfo {pages} {11169} (\bibinfo {year} {1976})}\BibitemShut {NoStop}%
\bibitem [{\citenamefont {Dronskowski}\ and\ \citenamefont {Bloechl}(1993)}]{Richard93p8617}%
  \BibitemOpen
  \bibfield  {author} {\bibinfo {author} {\bibfnamefont {R.}~\bibnamefont {Dronskowski}}\ and\ \bibinfo {author} {\bibfnamefont {P.~E.}\ \bibnamefont {Bloechl}},\ }\bibfield  {title} {\bibinfo {title} {Crystal orbital hamilton populations ({COHP}): energy-resolved visualization of chemical bonding in solids based on density-functional calculations},\ }\href {https://doi.org/10.1021/j100135a014} {\bibfield  {journal} {\bibinfo  {journal} {J. Phys. Chem.}\ }\textbf {\bibinfo {volume} {97}},\ \bibinfo {pages} {8617} (\bibinfo {year} {1993})}\BibitemShut {NoStop}%
\bibitem [{\citenamefont {Deringer}\ \emph {et~al.}(2011)\citenamefont {Deringer}, \citenamefont {Tchougréeff},\ and\ \citenamefont {Dronskowski}}]{Volker11p5461}%
  \BibitemOpen
  \bibfield  {author} {\bibinfo {author} {\bibfnamefont {V.~L.}\ \bibnamefont {Deringer}}, \bibinfo {author} {\bibfnamefont {A.~L.}\ \bibnamefont {Tchougréeff}},\ and\ \bibinfo {author} {\bibfnamefont {R.}~\bibnamefont {Dronskowski}},\ }\bibfield  {title} {\bibinfo {title} {Crystal orbital hamilton population ({COHP}) analysis as projected from plane-wave basis sets},\ }\href {https://doi.org/10.1021/jp202489s} {\bibfield  {journal} {\bibinfo  {journal} {J. Phys. Chem. A.}\ }\textbf {\bibinfo {volume} {115}},\ \bibinfo {pages} {5461} (\bibinfo {year} {2011})}\BibitemShut {NoStop}%
\bibitem [{\citenamefont {Maintz}\ \emph {et~al.}(2016)\citenamefont {Maintz}, \citenamefont {Deringer}, \citenamefont {Tchougréeff},\ and\ \citenamefont {Dronskowski}}]{Stefan16p1030}%
  \BibitemOpen
  \bibfield  {author} {\bibinfo {author} {\bibfnamefont {S.}~\bibnamefont {Maintz}}, \bibinfo {author} {\bibfnamefont {V.~L.}\ \bibnamefont {Deringer}}, \bibinfo {author} {\bibfnamefont {A.~L.}\ \bibnamefont {Tchougréeff}},\ and\ \bibinfo {author} {\bibfnamefont {R.}~\bibnamefont {Dronskowski}},\ }\bibfield  {title} {\bibinfo {title} {{LOBSTER}: A tool to extract chemical bonding from plane‐wave based {DFT}},\ }\href {https://doi.org/10.1002/jcc.24300} {\bibfield  {journal} {\bibinfo  {journal} {J. Comput. Chem.}\ }\textbf {\bibinfo {volume} {37}},\ \bibinfo {pages} {1030} (\bibinfo {year} {2016})}\BibitemShut {NoStop}%
\bibitem [{\citenamefont {Plimpton}(1995)}]{Plimpton95p1}%
  \BibitemOpen
  \bibfield  {author} {\bibinfo {author} {\bibfnamefont {S.}~\bibnamefont {Plimpton}},\ }\bibfield  {title} {\bibinfo {title} {Fast parallel algorithms for short-range molecular synamics},\ }\href {https://doi.org/10.1006/jcph.1995.1039} {\bibfield  {journal} {\bibinfo  {journal} {J. Comput. Phys.}\ }\textbf {\bibinfo {volume} {117}},\ \bibinfo {pages} {1} (\bibinfo {year} {1995})}\BibitemShut {NoStop}%
\end{thebibliography}%

\clearpage
{\bf{Acknowledgments}} The computational resource is provided by Westlake HPC Center. 

{\bf{Author Contributions}} Y.T. and S.L.~conceived the idea and designed the project. X.C. performed the simulations and analyzed the data. X.C., Y.T., and S.L.~wrote the manuscript. 
All authors discussed the results and commented the manuscript.

{\bf{Competing Interests}} The authors declare no competing financial or non-financial interests.

{\bf{Data Availability}} The data that support the findings of this study are included in this article and are available from the corresponding author upon reasonable request.

\end{document}